\documentclass[a4paper,usenatbib]{mnras}

\usepackage{amsmath}
\usepackage[switch]{lineno}
\usepackage{graphicx}

\title[Monitoring VHE transients]{Detection of VHE gamma-ray transients with monitoring facilities}

\author[G. La Mura et al.]{G. La Mura\thanks{\textit{E-mail}: glamura@lip.pt},$^1$
 G. Chiaro,$^2$
 R. Concei\c{c}\~ao,$^{1,3}$
 A. De Angelis,$^{4,5,6}$
 M. Pimenta,$^{1,3}$ \newauthor
 B. Tom\'e$^{1,3}$\\
 $^{1}$Laborat\'orio de Instrumenta\c{c}\~ao e F\'{i}sica Experimental de Part\'{i}culas (LIP), Av. Prof. Gama Pinto 2, 1649-003 Lisboa, Portugal\\
 $^{2}$Istituto di Astrofisica Spaziale e Fisica cosmica - INAF , Via A. Corti 12, 20133 Milano, Italia\\
 $^{3}$Instituto Superior T\'ecnico (IST), Av. Rovisco Pais 1, 1049-001 Lisboa, Porugal\\
 $^{4}$Dipartimento di Fisica e Astronomia - Universit\`a di Padova, Via Marzolo 8, 35131 Padova, Italia\\
 $^{5}$Dipartimento di scienze matematiche, informatiche e fisiche - Universit\`a degli Studi di Udine, Via Palladio 8, 33100 Udine, Italia\\
 $^{6}$Istituto Nazionale di Fisica Nucleare sez. Padova (INFN), Via Marzolo 8, 35131 Padova, Italia}

\date{Received 2020 June 18. Accepted 2020 July 17; in original form 2020 January 13}
\pubyear{2020}

\begin{document}
\newcommand{\de}{\mathrm d}
\maketitle
\begin{abstract}
  The observation of Very High Energy $\gamma$\ rays (VHE, $E > 100\,$GeV) led us to the identification of extremely energetic processes and particle acceleration sites both within our Galaxy and beyond. We expect that VHE facilities, like CTA, will explore these sources with an unprecedented level of detail. However, the transient and unpredictable nature of many important processes requires the development of proper monitoring strategies, to observe them. With this study, we estimate the properties of VHE transients that can be effectively detected by monitoring facilities. We use data collected by the {\it Fermi}-LAT instrument, during its monitoring campaign, to select events that are likely associated with VHE emission. We use this sample to estimate the frequency, the luminosity and the time-scales of different transients, focusing on blazar flares and Gamma Ray Bursts (GRBs). We discuss how the balance between Field of View, sensitivity and duty cycle of an observatory affects the likelihood to detect transients that occur at the inferred rates and we conclude describing the contribution that current and near-future monitoring facilities can bring to the identification and study of VHE transient emission.
\end{abstract}


\section{Introduction}
The recent detection of a Gamma-Ray Burst (GRB) associated with a Gravitational Wave event \citep[GW170817,][]{Abbott17a, Abbott17b} and a flaring blazar consistent with the direction of an ultra-relativistic neutrino \citep[TXS~0506+056,][]{IceCube18} demonstrated the importance of $\gamma$-ray monitoring in the identification of multi-messenger events. In both cases, the existence of an electromagnetic counterpart to the signals was first identified by $\gamma$-ray instruments, namely the Gamma-ray Burst Monitor \citep[GBM,][]{GBMpaper}, for the neutron star merger originating GW170817, and the Large Area Telescope \citep[LAT,][]{LATpaper}, for the blazar flare associated with IceCube-170922A, the two instruments carried by the \textit{Fermi Gamma-ray Space Telescope}. Follow-up observations executed in different wavelengths led to the clarification of other important characteristics, such as the identification of the GRB counterpart as a \textit{kilonova} \citep[e.g.][]{Cowperthwaite17}, as well as the redshift and, hence, the distance and the luminosity of the processes \citep{Blanchard17, Paiano18}. In particular, the blazar TXS~0506+056 has been detected at Very High Energy (VHE, $E \geq 100\,$GeV) for the first time by MAGIC, shortly after the Fermi outburst \citep{Mirzoyan17}.

Detecting VHE photons from sources like blazars and GRBs has several important implications. On one hand, the emission of VHE radiation in close connection with the production of ultra-relativistic particles is a strong hint towards the role of these sources as cosmic particle accelerators. On the other, the interaction of $\gamma$ rays with the lower energy photon field, forming the Extragalactic Background Light (EBL), is an extremely powerful tool to constrain the effects of star formation and active galactic nuclei (AGN) in the process of cosmological evolution. During its long monitoring campaign, the {\it Fermi}-LAT telescope has firmly identified AGNs - blazars in particular - and GRBs as the most powerful extragalactic sources of photons above $10\,$GeV \citep{3FHLpaper, Ajello19}. Some AGNs are known to have energy spectra that extend up to several TeV, while the combination of observed GRB spectra with the identification of reliable counterparts at measurable redshifts suggested that GRBs could be intrinsically able to produce photons well above $E = 100\,$GeV. This expectation would be eventually confirmed by the MAGIC and H.E.S.S. observations of some GRBs \citep{Mirzoyan19, GRBpaper2, GRBpaper}.

In recent times, our ability to observe VHE sources greatly improved, thanks to the construction of large ground-based observatories using either the Imaging Atmospheric Cherenkov Telescope approach \citep[IACT, like H.E.S.S., VERITAS, and MAGIC,][]{HESSpaper, VERITASpaper, MAGICpaper} or the Extensive Air Shower detector array \citep[EAS, such as HAWC and LHAASO,][]{HAWCpaper, LHAASOpaper}. Due to the strong implications of VHE observations on the physics of relativistic jets and light propagation through the Universe, great efforts are currently ongoing to improve the characteristics of these observatories. We can expect that the upcoming Cherenkov Telescope Array \citep[CTA,][]{CTApaper} will achieve outstanding performances in this field. Although CTA will be able to observe the VHE sky at an unprecedented level of detail, its ability to perform regular monitoring of sources, to map vast regions of the sky and to promptly respond to fast transients will be strictly limited by its narrow field of view and its duty cycle. While dedicated monitoring programs, such as the one carried out by FACT \citep{FACTpaper}, can cover a list of selected targets and, possibly, provide transient follow-up capabilities, the necessary observational requirements imply unavoidable gaps in the data flow. On the other hand, the presence of survey instruments that continuously scan wide areas of the sky, like Fermi, is essential to study transient phenomena. Unfortunately, the maintenance of long term survey missions in space is hard to attain and it comes at the cost of reduced efficiency in the VHE domain. It has been proposed that the observational coverage and continuity problems can be overtaken using the wide Field of View (FoV) of EAS facilities. In this work, we use $\gamma$-ray monitoring data to study the distribution of transients that are more likely to be associated with VHE emission. We analyze the properties of these events and we compare them with the performance of instruments used to monitor or follow-up transients. The paper is structured as follows: in \S2, we use {\it Fermi}-LAT data to study the distribution of events that are known or predicted to produce VHE emission at a detectable level for CTA, describing our selection, assumptions and implications; in \S3, we discuss the possibilities that EAS arrays have to contribute to VHE transient monitoring; in \S4 we compare different instruments and approaches and, finally, in \S5 we give our conclusions.

\section{Selection of VHE gamma-ray transients}
VHE transients are a fundamental probe of multi-messenger astrophysics and particle physics, but, at present, very little is known about the rate and the properties of these events. Since the performance of VHE instruments will have a critical impact on our ability to explore these events, we carry out a systematic study to assess how monitoring facilities can contribute to their investigation, by taking into account the source positions in the sky, their redshift, and the time-scales with which they can be detected. We develop this analysis in a sequence of steps that, starting from the available data, leads us to infer what instrument characteristics are required and which performances can be achieved. As a starting point, we use Fermi-LAT monitoring data to select a sample of transients that led to HE emission and are most likely connected with VHE activity. We use the fact that some of these transients are associated with sources that have been already detected at VHE, suggesting that their spectra are not subject to severe cut-offs. Finally, we compare the selected data and our assumptions with the performance of different observing facilities, demonstrating that instruments with optimized monitoring capabilities can both act as effective alert systems and explore a spectral window that would be otherwise very difficult to monitor.

\subsection{The spectrum of sources}
We want to determine the distribution and the properties of the most important extragalactic VHE sources. To achieve this purpose, we need to use a model that can be compared with observational data. In general, we may express the VHE photon spectrum observed from an astrophysical source at redshift $z$ in the form of:
\begin{eqnarray}
 \frac{\de N(E)}{\de E} &=& N_0 \left( \frac{E}{E_0} \right)^{-[\alpha + \beta \mathrm{log}\,(E/E_0)]} \mathrm{e}^{-[\tau_E(z) + E / E_{\rm c.o.}]} \nonumber \\
 & & \mathrm{[GeV^{-1}\, cm^{-2}\, s^{-1}]} \label{eqSEDform},
\end{eqnarray}
where $\alpha$ is the photon index ($\alpha \geq 1.5$ for most astrophysical sources), $\beta$ is the curvature parameter ($\beta = 0$ for a pure power-law spectrum), $E_0$ is a scaling energy, $E_{\rm c.o.}$ is the cut-off energy, and $\tau_E(z)$ is the Universe opacity at energy $E$ as a function of redshift.

Eq.~(\ref{eqSEDform}) implies that the number of photons available at energies $E \gg E_0$ can quickly become lower than 1 photon every few square metres. This means that instruments with small collecting areas, like those carried by satellites, require long observing times to characterize the high energy spectra of sources and they may not be able to effectively detect the highest energy emission of a short time-scale event. Thanks to the long duration of the {\it Fermi}-LAT monitoring campaign, however, we are able to select a sample of sources, among the ones which were detected at high energy (HE, $E > 10\,$GeV) and to study their flaring activity.

\subsection{AGN flares}
Since most EBL models predict that the Universe optical depth for $\gamma$-ray photons with $E \simeq 1\,$TeV is $\tau_{\rm 1 TeV} \geq 1$ already at $z \approx 0.1$ \citep{Desai19}, it turns out that the study of photons with $E < 1\,$TeV is critical to constrain EBL properties. However, while we know a handful of extragalactic sources that feature a persistent TeV emission, the vast majority of VHE sources are characterized by generally unpredictable activity.

To study the distribution of VHE flaring sources, we undertook an analysis aimed at characterizing the frequency, the duration, and the possible spectral features of VHE AGN flares. We used the $2^{\rm nd}$ {\it Fermi}-LAT All-Sky Variability Analysis \citep[2FAV,][]{2FAVApaper}\footnote{\texttt{https://fermi.gsfc.nasa.gov/ssc/data/access/lat/FAVA/}} as a starting point to select a uniform sample of $\gamma$-ray flares. Since the second version of the catalog includes $7.4\,$yr of variability analysis, with tools designed to inspect the $\gamma$-ray light curves of specific sky areas and the possibility to extract preliminary spectral fits to the soft ($100$-$800\,$MeV) and the hard ($0.8$-$300\,$GeV) LAT bands, we are able to search the LAT data for the brightest outbursts detected on a weekly time-scale. We therefore performed a selection of AGNs that have been associated with $\gamma$-ray flares and were detected with an energy flux larger than $10^{-12}\, \mathrm{erg\, cm^{-2}\, s^{-1}}$ above $10\,$GeV in the Third Catalog of Hard {\it Fermi}-LAT sources \citep[3FHL,][]{3FHLpaper}. This selection led to the identification of 160 sources, which have been associated with 2367 $\gamma$-ray flares, detected in the 2FAV hard band.

Although using the {\it Fermi}-LAT variability analysis to study the properties of flares has some limitations, we can identify examples of flares that triggered specific follow-up analysis. Comparing 2FAV selected flares with almost simultaneous analyses of data suggests that the power-law fits to the 2FAV hard band give reasonable representations of the spectral energy distribution (SED). In general, flares are associated with harder spectral indexes than those that characterize the average SEDs of the $8\,$yr data set of the Fourth {\it Fermi}-LAT Gamma-ray Catalogue \citep[4FGL,][]{4FGLpaper}. While this procedure can effectively identify the events with strongest HE activity, the extrapolation of spectra towards the VHE domain is subject to some degree of uncertainty, because sources may exhibit intrinsic spectral cut-offs. These, however, are not expected to be found below an energy of approximately 300 GeV. Indeed, some objects associated with our flare sample are detected by Fermi-LAT above $100\,$GeV, although several years of monitoring were necessary to collect the required statistics. In addition, many of the sources that exhibited the brightest flares have been detected at VHE by IACT facilities \citep{TeVCatpaper}, suggesting that intrinsic spectral cut-offs are unlikely to occur below $500\,$GeV. Therefore, in general, we do not expect severe cut-offs in flaring states, where, on the contrary, many works point to the existence of possible additional emission components \citep{Abdalla17, Prince17, Zacharias19}, while it is well established that the blazar TXS~0506+056 was first detected in the VHE domain after being identified as a LAT flaring blazar consistent with an IceCube neutrino event \citep{Abeysekara18}.

\begin{figure}
  \begin{center}
    \includegraphics[width=0.45\textwidth]{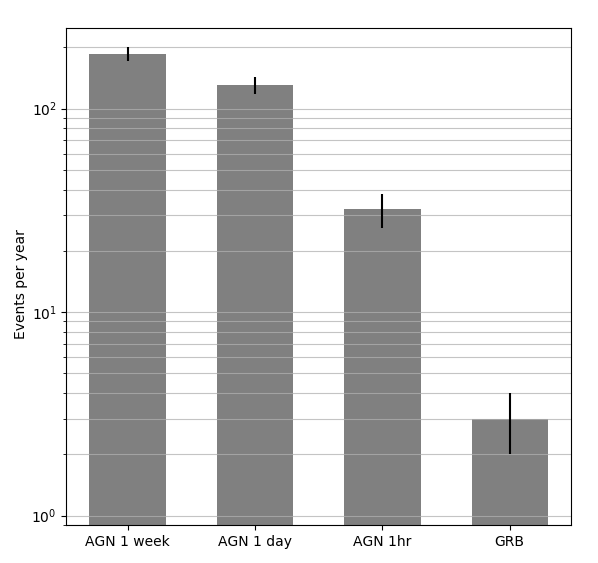}
    \caption{The estimated frequency of VHE transients associated with AGN flares of different duration and with GRBs. The events taken into account are limited to the ones which are expected to be bright enough to be detected by CTA South if they were located within $20^\mathrm{o}$ from zenith, and they could be observed for $1\,$hr. \label{figTransientFreq}}
  \end{center}
\end{figure}
The selected data describe observations under the effect of the EBL opacity. Therefore, we can use the spectral parameters inferred from the variability analysis to estimate the flux produced by flaring sources in the VHE domain. To do this, we apply the EBL absorption model of \citet{Dominguez11} to the extrapolation of Eq.~\ref{eqSEDform} to $E > 100\,$GeV and we use the resulting fluxes to select flaring events, which are estimated to be brighter than $4.40 \cdot 10^{-12}\, {\rm erg\, cm^{-2}\, s^{-1}}$ between $100\,$GeV and $200\,$GeV, roughly corresponding to the limiting flux for a $5\sigma$\ detection with CTA South (CTA-S) in $1\,$hr of observation. This further limitation leads to the selection of 1374 flares that were associated with AGNs in $7.4\,$yr of monitoring, and, thus, to an average flare rate of $187$ events per year that increased the flux at a detectable level for the time-scale of $1$ week. Following this method we obtain a list of transients that CTA-S would be able to detect if every event lasted exactly one week, with a flux constantly equal to its average value, as computed on a fixed time binning, and provided that the target could be observed for $1\,$hr during the transient. In a more realistic scenario, the flaring activity is not limited to a smooth increase up to a constant luminosity, but it can include shorter features that increase the flux of factors of $3$ over one day, and, more rarely, reach peaks of nearly $10$ times the average for a few hours. Adopting these assumptions on the internal evolution of our sample of flares and assuming a Poissonian distribution to account for the underlying uncertainty, we can estimate that $963$ total events may have led to a transient that would be brighter than our limiting flux for a few days (corresponding to an average rate of $130 \pm 12$ per year) and $237$ in one that has been detectable for a few hours (resulting in averagely $32 \pm 6$ per year). Considering the duration of the monitoring campaign, we can therefore expect that VHE transients occur with the average rates shown in Fig.~\ref{figTransientFreq}.

\begin{figure}
 \centering
 \includegraphics[width=0.48\textwidth]{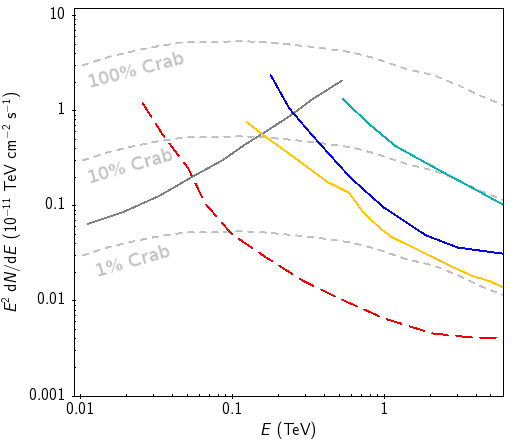}
 \caption{Differential sensitivity to a point-like source for HAWC (cyan continuous line), LHAASO (blue line), and SWGO (yellow line) as compared with the {\it Fermi}-LAT Pass 8 sensitivity (dark grey line) computed on 1 year of observations. For comparison, the plot also shows different fractions of the Crab Nebula flux spectrum (short dashed light grey curves), as well as the sensitivity achieved by CTA-S in $50\,$hours of observation (long dashed red line). \label{figSensCurves}}
\end{figure}
\subsection{Gamma-Ray Bursts}
At present, the data that we possess on the VHE properties of GRBs are still very scarce, and any attempt to model them are subject to large uncertainties, due to the many important free parameters that affect the predicted luminosities and spectra \citep{Galli08, Bernardini19}. If we are interested in an estimate of the rate of VHE events, we can start from the observation that GRBs detected with a maximum photon energy above $100\,$GeV are brighter towards the lower energy limit. As a consequence, we expect that GRBs with a significant VHE emission should in principle be detected by {\it Fermi}-LAT in the HE domain, if properly pointed.

According to the data presented in the second catalog of LAT detected GRBs \citep{Ajello19}, in 10 years of operation, the LAT has been able to detect 169 GRBs with photons above $100\,$MeV, while only 15 events were associated with photons detected above $10\,$GeV. By normalizing the detection rates with respect to the LAT effective area inferred from the P8R3\_TRANSIENT\_V2 instrument response function (IRF) and assuming an isotropic distribution of GRBs, we can relate the different detection rates with the probability that a specific GRB spectrum extends above a critical energy. If we denote as $N_{GRB}(E)$ the cumulative number of events per year that emit photons up to energy $E$ and we assume an isotropic GRB distribution, we can reproduce the observed GRB rate from a power-law distribution in the form of:
\begin{equation}
 N(E) = 291 \left( \frac{E}{100\, \mathrm{MeV}} \right)^{-1.5}. \label{eqGRBpop}
\end{equation}
Clearly, a simple extrapolation of the observed LAT trend is still a very limited approximation of the real GRB properties, because the reduced sensitivity of the LAT above $100\,$GeV and the extreme rarity of the most energetic photons can lead to a poor sampling of the actual high energy properties. However, the emission of VHE photons requires favourable energetic conditions in the source and, since the EBL opacity limits their propagation within a relatively small horizon, we can expect that a rate of at least $3\, (\pm 2)$ GRBs per year may be associated with detectable VHE emission (see Fig.~\ref{figTransientFreq}).

\begin{figure*}
 \centering
 \includegraphics[width=0.95\textwidth]{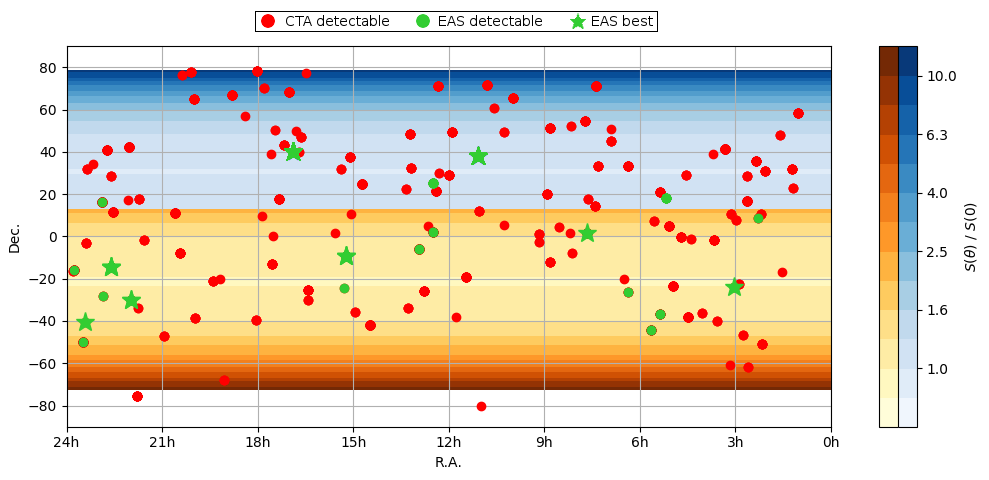}
 \caption{Distribution of VHE flaring blazars, compared with the sky coverage of LHAASO (blue shaded region) and SWGO (orange shaded region) within a FoV of $50^{\mathrm{o}}$ from zenith, assuming an SWGO site with latitude $23^{\mathrm{o}}$S. The colour shading represents the sensitivity degradation with respect to a source that culminates at the zenith. The sensitivity comparison in the overlapping region is computed at $300\,$GeV and only the instrument with the best estimated performance is plotted. The red dots represent objects for which {\it Fermi}-LAT detected flares that would be bright enough to be observed by CTA in $1\,$hr, but are too faint for the computed EAS sensitivities. The green dots are sources whose flares are comparable to the sensitivity of the relevant monitoring instrument, while the stars represent objects whose flares are either already detected by HAWC, or their {\it Fermi}-LAT spectral fit is significantly brighter than the predicted EAS sensitivity limit scaled down to the duration of the flare. Every source can produce more than one flare. \label{figFlaresMap}}
\end{figure*}
\section{VHE transient event monitoring}
With our study of {\it Fermi}-LAT $\gamma$-ray transients, we estimated that several hundreds transient events can represent high priority targets for sub-TeV VHE investigations every year. Their unpredictable nature and the possibility that a significant fraction of triggers issued by lower energy monitoring instruments may have little or no VHE emission, however, poses the question to explore what instrument performance is required to identify the most relevant ones. While CTA is obviously expected to provide the best sensitivity in this energy range, the performance of monitoring instruments on short time-scales is still very limited. {\it Fermi}-LAT has low sensitivity to VHE photons, while the largest ground based facilities of HAWC and LHAASO are located in the Northern hemisphere and, therefore, do not provide full-sky coverage.

VHE particles - either cosmic rays or photons - that interact with the atmosphere generate a shower of relativistic secondary products, including charged particles and Cherenkov radiation, which develops downward, approximately in the direction of the incoming primary. These showers can be tracked by collecting the Cherenkov light and by detecting the secondary particles that reach the ground. Arrays of particle detectors, placed on the ground, but at high altitude, can infer the direction and the energy of atmospheric showers, provided that they are able to determine with sufficient accuracy the location, the arrival time and the energy carried by the secondary charged particles that reach them. For this work, we took the yellow curve in Fig.~\ref{figSensCurves}, as the reference curve for SWGO\footnote{www.swgo.org} \citep{SWGOpaper}. This curve was for the most part obtained through a straw-man model which assumed a facility based on water Cherenkov detectors (WCD) covering an area of $80\,000\,{\rm m}^2$ with an $80\%$  filling factor~\citep{SGSOpaper}. For the lower part, below 300 GeV, the most important energy region for this work, the assumed sensitivity is based on an end-to-end simulation \citep[performed with Geant4,][]{GEANT4}  of  a detector concept which combines a WCD with white reflective walls and a Resistive Plate Chamber (RPC), to effectively lower the energy threshold~\citep{LATTESpaper}. The success of this concept relies on the ability to trigger on low-energy secondary photons with the WCD while being able to measure time with a resolution better than $2\,$ns, crucial for a good geometry reconstruction. The instrument performances illustrated in Fig.~\ref{figSensCurves} are computed for a steady point source located at a zenith distance of $20^{\rm o}$ and under the assumption that it can be observed $6\,$hr per day. For wide FoV instruments, we need to correct this sensitivity according to the visibility of different sky areas.

\subsection{Sensitivity across the Field of View}
Although a wide FoV instrument can track simultaneously targets located in different sky regions, the corresponding sensitivity depends on the zenith distance of every source. Due to computational limitations, the dependence of the sensitivity with the shower inclination was obtained using the amount of shower electromagnetic energy at the ground as a proxy.  This allowed us to use the shower simulations generated with CORSIKA \citep{CORSIKA}, while skipping the full detector simulation and the application of the shower reconstruction analyses,  as described in detail in \citet{LATTESpaper}.  While this is a crude estimate,  we believe it is  a conservative approach. On the one hand, the trigger probability decreases with the increase of the zenith angle, as fewer particles (energy) reach the ground due to the atmosphere increasing thickness. On the other hand, the shower is more spread over the array, which eases the geometric reconstruction. This has an impact not only on the astrophysical source position determination but also effectively reduces the hadronic background.

Since the Earth's daily rotation induces a transit that changes the fraction of time that a source spends at a specific zenith distance $\theta$, depending on the latitude of the observing site and the source's declination, we derived the fraction of time $\Delta t(\theta)$ that every point in the sky spends at a given zenith distance $\theta$, according to its declination. If an instrument operates for an observing time $\Delta t$, under conditions that change with time, like $\theta$\ for a transiting source, the sensitivity of the observation scales as:
\begin{equation}
 S(\Delta t) = \left[ \frac{1}{T_0} \sum_i \frac{\Delta t_i}{S_i^2(T_0)} \right]^{-1/2} \label{eqEffSens}
\end{equation}
where $\Delta t_i$ are the amounts of time during which we can consider the instrument to have a regular performance, while $T_0$ and $S_i(T_0)$ represent, respectively, the standard time interval, on which the sensitivity is computed, and the sensitivity under constant observing conditions over such interval.

To derive the effective sensitivities that would result by the combined operation of LHAASO and SWGO observatories, we used the corresponding sensitivity curves, computed for one year of observations, assuming that SWGO is located at a latitude of $23^\mathrm{o}$S.\footnote{This is the latitude of the ALMA site, one of the possible candidate sites for such experiment.} We subdivided the sky in $2^{\mathrm{o}}$ wide strips of constant declination, and we computed the fraction of time that a point belonging to each one of these strips spends at a zenith distance $\theta \leq 50^{\mathrm{o}}$. We used the sensitivities computed as a function of $\theta$\ for an energy of $300\,$GeV \citep{LATTESpaper} to obtain the sensitivity degradation resulting when the zenith distance increases from $\theta = 0^{\mathrm{o}}$ to $\theta = 50^{\mathrm{o}}$ in steps of $5^{\mathrm{o}}$. Finally, we applied Eq.~(\ref{eqEffSens}) to extract the sensitivity that an observation can effectively achieve in the whole FoV, thus deriving an estimate of the limiting flux to detect a target within the monitored area. The result of this calculation, carried out for the SWGO and LHAASO experiments, is shown in Fig.~\ref{figFlaresMap}, together with the distribution of VHE flaring AGNs.

\begin{figure*}
 \centering
 \includegraphics[width=0.48\textwidth]{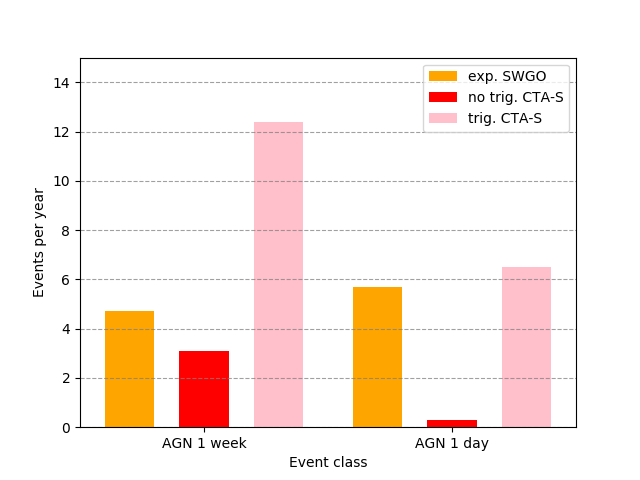}
 \includegraphics[width=0.48\textwidth]{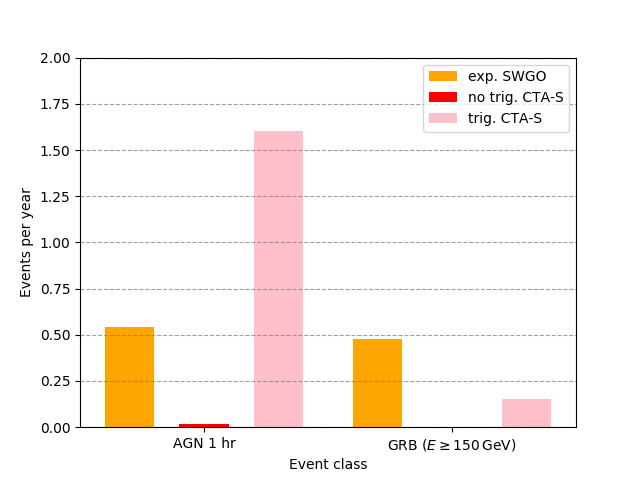}
 \caption{Histograms illustrating the predicted detection rates of VHE transients after $1\,$yr of observations with SWGO and CTA South. The color bars represent the expected rates at which VHE transients can be detected by SWGO (yellow), CTA-S without a trigger (red), and CTA-S assuming a perfect monitoring program that issues triggers for all VHE transients (pink). The analysis takes into account the instrument sensitivities, duty cycles and FoVs and divides the transients in long events, for which the transit over the observing site is granted (left panel) and short events, where the additional possibility that the transient is in an unobservable portion of the sky is taken into account (right panel). To estimate the CTA serendipitous detection rate, we assume that the area scanned by CTA increases with the duration of the flare as result of the combination of different pointings. \label{figProbHist}}
\end{figure*}
\section{Comparison between different experiments}
Taking as reference the performance of CTA-S,\footnote{\texttt{https://www.cta-observatory.org/science/cta-performance/}} our analysis of the {\it Fermi}-LAT monitoring data led to the identification of 160 blazars and of a fraction of GRBs distributed across the whole sky, whose light-curves show transient VHE emission that is strong enough for detection in $1\,$hr of observation. Since blazars are sources of potential VHE activity for which we know the position in the sky, we can compare their emission with the predicted sensitivity of monitoring instruments that regularly scan the sky, such as LHAASO and SWGO, as it is illustrated in Fig.~\ref{figFlaresMap}. The shaded areas represent the FoV covered within $50^{\mathrm{o}}$ from zenith, with a colour scale that illustrates the sensitivity degradation towards different sky regions, with respect to a target that culminates at the zenith. Only the instrument with the best estimated sensitivity is plotted in the overlapping region.

If we apply an extrapolation of Eq.~(\ref{eqSEDform}) to the spectra of the HE flares identified in our sample, taking into account the effects of EBL \citep{Dominguez11, Kudoda17}, we can compare the predicted VHE fluxes with the chances that a monitoring instrument has to detect a specific transient. With the assumed instrument performance, it turns out that $21$ blazars out of $160$, listed in Table~\ref{tabAGNSample}, showed a flaring activity that is comparable to the detection threshold of monitoring facilities (green symbols). Eight of these sources, marked as stars in Fig.~\ref{figFlaresMap} and listed in bold-face in Table~\ref{tabAGNSample}, have a predicted spectrum that is significantly brighter than the limiting flux expected for the instrument covering their location in the sky, or, in the case of MRK~421 and MRK~501, are already detected by HAWC, while six of them appear in the {\it TeVCat} list of objects detected in VHE domain. The remaining sources, instead, show an estimated flaring activity that is too faint for detection by a monitoring instrument. All of these objects, however, remain of potential interest, because most of them have been associated with multiple flares and the use of their averaged spectral properties can still smear out the possible existence of sharp peaks in their light-curve at shorter time-scales.

Indeed, the importance to have efficient monitoring instruments is connected with the chances to detect and track flaring activity. CTA, with an observing budget of $1500\,$hr per year per site, corresponding to a duty cycle smaller than $20$ per cent \citep{Actis11}, and in the assumption that it can collect observational data on average for $5\,$hr per night, covering a FoV with an angular radius of $2^\mathrm{o}$, has practically null chances to be in place, when a flare occurs. Assuming that CTA follows a scheduled observing plan (i.e., not taking into account Target of Opportunity observations), the probability to detect a transient in a random position of the observable sky is proportional to the rate of the transient and to the ratio between the area covered by CTA observations and the total visible sky (approximately $2\,$sr, below $1\,$TeV). Recalling that our estimate of transient rates are based on the events which are bright enough to be detected in $1\,$hr of observation and assuming that CTA exposure have no significant overlap above $1\,$hr, the scanned area increases with the duration of the flare, resulting in a higher detection chance for longer transients. Clearly, the detection probability can be even higher, if CTA is alerted about the occurrence of a transient, with accurate positional information. At present, however, the available instruments are not able to filter which low energy transients will be actually associated with VHE emission and following all the possible triggers would probably result in a high rate of false alarms.

\begin{table*}
\caption{List of AGNs that have been detected by {\it Fermi}-LAT with an energy flux larger than $10^{-12}\, \mathrm{erg\, cm^{-2}\, s^{-1}}$ above $10\,$GeV and associated with flaring activity above the sensitivity limit of monitoring facilities. The columns report the name of the AGN, its 3FHL association, the sky coordinates (Right Ascension and Declination, J2000), the redshift, the number of associated flares in $7.4\,$years, a flag stating whether the source is listed in \textit{TeVCat}, and the energy flux obtained from a power-law fit to the 3FHL data. We mark with bold-faced names those whose flares are significantly stronger than the corresponding detection limit (marked with stars in Fig.~\ref{figFlaresMap}). \label{tabAGNSample}}
\begin{footnotesize}
\begin{tabular}{lccccccc}
\hline
\hline
{\bf Name} & {\bf 3FHL source} & {\bf R.A.} & {\bf Dec.} & $\boldsymbol{z}$ & $\boldsymbol{N}_{flares}$ & {\bf TeVCat} & {\bf 3FHL en. flux}\\
 & & hh:mm:ss & dd:mm:ss & & & & $10^{-12}\, \mathrm{erg\, cm^{-2}\, s^{-1}}$ \\
\hline
TXS 0214+083 & J0217.1+0836 & $02:17:17.12$ & $+08:37:03.89$ & $0.085$ & 2 & N & $2.248 \pm 0.789$ \\
{\bf PKS 0301--243} & J0303.4--2407 & $03:03:26.50$ & $-24:07:11.42$ & $0.260$ & 3 & Y & $36.892 \pm 4.831$ \\
PKS 0507+17 & J0510.0+1800 & $05:10:02.37$ & $+18:00:41.58$ & $0.416$ & 9 & N & $5.191 \pm 1.034$ \\
PKS 0521--365 & J0523.0--3627 & $05:22:57.98$ & $-36:27:30.85$ & $0.055$ & 11 & N & $5.520 \pm 1.610$ \\
PKS 0537--441 & J0538.8--4405 & $05:38:50.36$ & $-44:05:08.94$ & $0.892$ & 91 & N & $37.847 \pm 2.540$ \\
PMN J0622--2605 & J0622.4--2606 & $06:22:22.06$ & $-26:05:44.64$ & $0.414$ & 3 & N & $11.495 \pm 2.740$ \\
{\bf PKS 0736+0174} & J0739.3+0137 & $07:39:18.03$ & $+01:37:04.62$ & $0.191$ & 33 & Y & $2.986 \pm 1.039$ \\
{\bf MRK 421} & J1104.4+3812 & $11:04:27.31$ & $+38:12:31.80$ & $0.031$ & 9 & Y & $437.000 \pm 18.600$ \\
3C 273 & J1229.2+0201 & $12:29:06.70$ & $+02:03:08.60$ & $0.158$ & 66 & N & $1.309 \pm 0.496$ \\
ON 246 & J1230.2+2517 & $12:30:14.09$ & $+25:18:07.14$ & $0.135$ & 18 & Y & $8.428 \pm 1.621$ \\
3C 279 & J1256.1--0547 & $12:56:11.17$ & $-05:47:21.53$ & $0.536$ & 79 & Y & $18.588 \pm 2.200$ \\
{\bf PKS 1510--089} & J1512.8--0906 & $15:12:50.53$ & $-09:05:59.83$ & $0.360$ & 187 & Y & $35.061 \pm 3.116$ \\
Ap Librae & J1517.6--2422 & $15:17:41.81$ & $-24:22:19.48$ & $0.049$ & 1 & Y & $20.327 \pm 3.354$ \\
{\bf MRK 501} & J1653.8+3945 & $16:53:52.22$ & $+39:45:36.61$ & $0.033$ & 2 & Y & $156.000 \pm 11.000$ \\
{\bf PKS 2155--304} & J2158.8--3013 & $21:58:52.06$ & $-30:13:32.12$ & $0.116$ & 3 & Y & $132.957 \pm 8.901$ \\
{\bf PKS 2233--148} & J2236.5--1433 & $22:36:34.09$ & $-14:33:22.19$ & $0.325$ & 17 & N & $9.309 \pm 1.620$ \\
PMN J2250--2806 & J2250.7--2806 & $22:50:44.49$ & $-28:06:39.32$ & $0.525$ & 6 & N & $4.811 \pm 1.104$ \\
3C 454.3 & J2253.9+1608 & $22:53:57.75$ & $+16:08:53.56$ & $0.859$ & 301 & N & $32.758 \pm 2.202$ \\
{\bf 1ES 2322--40.9} & J2324.7--4040 & $23:24:44.67$ & $-40:40:49.44$ & $0.174$ & 1 & N & $9.439 \pm 2.197$ \\
PKS 2326--502 & J2329.2--4955 & $23:29:20.88$ & $-49:55:40.64$ & $0.518$ & 76 & N & $7.931 \pm 0.993$ \\
PMN J2345--1555 & J2345.1--1554 & $23:45:12.56$ & $-15:55:07.83$ & $0.621$ & 34 & N & $2.027 \pm 0.656$ \\
\hline
\end{tabular}
\end{footnotesize}
\end{table*}
Conversely, if we focus our attention to the sky region that will be covered by CTA-S, because of its better performance, we can illustrate the role of a monitoring facility, by looking at the predicted detection rates illustrated in Fig.~\ref{figProbHist}. If we use the results of our analysis to infer the average rate of transients with detectable VHE emission, applying visibility and duty cycle constraints, we can expect that CTA will be able to detect dozens of transients per year with just $1\,$hr of observation, if every VHE event were associated with an accurate positional trigger. The number of potentially detectable sources is naturally expected to decrease for events of shorter time-scale and to drop practically at zero, in absence of a trigger. At present, the largest set of monitoring data in the HE domain is the one provided by the {\it Fermi}-LAT, but even this information can only be accessed after a time lag of approximately $6\,$hr, due to data down-link and processing requirements. As a result, the CTA ability to detect transients, without a monitoring facility, can be seriously affected. The existence of a monitoring instrument with the quoted SWGO performance would recover a significant fraction of the longest duration triggers and, possibly, achieve even better results on the shortest ones. This strategy, therefore, would represent a viable system to trigger CTA observations on the most relevant VHE transient events and also to collect independent data, extending our ability to monitor VHE activity towards sources that lie at $z \approx 0.3$ and possibly beyond.

\section{Conclusions}
The study of VHE transient phenomena is growing in importance, now that flaring activity in blazars has been shown to take part in the acceleration of ultra-energetic particles and that VHE photons have been firmly detected from GRBs. In this work we analyzed the results of the $\gamma$-ray monitoring campaign, carried out by the {\it Fermi}-LAT instrument, to identify the sources of flaring activity that are most likely associated with transient VHE emission. We estimated the expected fluxes of these VHE transients and we compared them with the predicted performance of CTA-S and of monitoring facilities like LHAASO and SWGO. We pointed out that a fraction of the selected sources is expected to have strong enough flaring activity to be tracked by EAS arrays that continuously scan wide portions of the sky. We show that instruments with optimized sub-TeV capabilities are able to provide relevant information on a substantial fraction of events that CTA could only detect if properly triggered. At present, most of the triggers are provided by satellites at relatively low energy, while the high energy information only comes with some delay (for instance, it takes approximately $6\,$hr before LAT data can be processed). EAS arrays, on the other hand, can directly trigger on the VHE band, with shorter response times. Given that EAS arrays can cover very large instrumented surfaces at a relatively low cost and that they have the further advantages to be maintainable and upgradable, with respect to a space mission, we point out that obtaining good sensitivities to the sub-TeV range of these detectors will bring to major improvements in the role of VHE transients as cosmic messengers.

\section*{Data availability}
There are no new data associated with this article.

\section*{Acknowledgements}
The authors would like to thank F. Longo, N. Omodei and the referee for useful discussion and suggestions.

This work was partly performed under project PTDC/FIS-PAR/29158/2017, Funda\c{c}\~ao para a Ci\^encia e Tecnologia. RC is grateful for the financial support by OE - Portugal, FCT, I. P., under DL57/2016/cP1330/cT0002.

The {\it Fermi}-LAT Collaboration acknowledges generous ongoing support from a number of agencies and institutes that have supported both the development and the operation of the LAT as well as scientific data analysis. These include the National Aeronautics and Space Administration and the Department of Energy in the United States, the Commissariat \`a l'Energie Atomique and the Centre National de la Recherche Scientifique / Institut National de Physique Nucl\'eaire et de Physique des Particules in France, the Agenzia Spaziale Italiana and the Istituto Nazionale di Fisica Nucleare in Italy, the Ministry of Education, Culture, Sports, Science and Technology (MEXT), High Energy Accelerator Research Organization (KEK) and Japan Aerospace Exploration Agency (JAXA) in Japan, and the K. A. Wallenberg Foundation, the Swedish Research Council and the Swedish National Space Agency in Sweden.

This research has made use of the CTA instrument response functions provided by the CTA Consortium and Observatory, see https://www.ctaobservatory.org/science/cta-performance/ (version prod3b-v2) for more details.

Additional support for science analysis during the operations phase is gratefully acknowledged from the Istituto Nazionale di Astrofisica in Italy and the Centre National d'\'Etudes Spatiales in France. This work performed in part under DOE Contract DEAC02-76SF00515.

\bibliographystyle{mnras}
\bibliography{ms_mnras}
\end{document}